\documentclass[]{raa}           
\usepackage{graphicx,times,picture}
\usepackage{natbib}
\usepackage{amssymb,amsmath}
\bibpunct{(}{)}{;}{a}{}{,}

\begin{document}

   \title{To understand the X-ray spectrum of anomalous X-ray pulsars and soft gamma-ray repeaters
}

 \volnopage{ {\bf 2013} Vol.\ {\bf X} No. {\bf XX}, 000--000}
   \setcounter{page}{1}

   \author{Yan-Jun Guo\inst{1}, Shi Dai \inst{1}, Zhao-Sheng Li \inst{1}, Yuan Liu \inst{2}, Hao Tong \inst{3}, Ren-Xin Xu \inst{1,4}
   }

   \institute{ School of Physics, Peking University, Beijing 100871, China; 
   {\it guoyj10@pku.edu.cn}\\
        \and
        Institute of High Energy Physics, Chinese Academy of Sciences, Beijing 100049, China
        \and
        Xinjiang Astronomical Observatory, Chinese Academy of Sciences, Urumqi 830011, China
        \and
        Kavli Institute for Astronomy and Astrophysics, Peking University, Beijing 100871, China
   }

   \date{Received~~2009 month day; accepted~~2009~~month day}

\abstract{Hard X-rays above 10 keV are detected from several anomalous X-ray pulsars (AXPs) and soft gamma-ray repeaters (SGRs), and different models have been proposed to explain the physical origin within the frame of either magnetar model or fallback disk system. 
   %
  Using data from {\it Suzaku} and INTEGRAL, we study the soft and hard X-ray spectra of four AXPs/SGRs, 1RXS J170849-400910, 1E 1547.0-5408,  SGR 1806-20 and SGR 0501+4516.
   It is found that the spectra could be well reproduced by bulk-motion Comptonization (BMC) process as was first suggested by Tr{\"u}mper et al., showing that the accretion scenario could be compatible with X-ray emission of AXPs/SGRs.
   %
   HXMT simulations for BMC model show that the spectra would have discrepancies from power-law, especially the cutoff at $\sim$ 200 keV.
   Thus future observations are promising to distinguish different models for the hard tail and may help us understand the nature of AXPs/SGRs.
}

   \authorrunning{Y.-J. Guo et al. }            
   \titlerunning{To understand the X-ray spectrum of AXPs and SGRs}  
   \maketitle

%
\section{Introduction}           

Since the discovery of radio pulsars in 1967, various kinds of pulsar-like objects have been observed, with diverse manifestations. 
Among them, anomalous X-ray pulsars (AXPs) and soft gamma-ray repeaters (SGRs) are peculiar kinds of sources~\citep{Mereghetti08}.
Their persistent X-ray luminosities are much higher than spin down energy, while no binary signature has been observed, thus ruling out both rotation and accretion in binary system as power source.
They also show recurrent bursts in the hard X-ray/soft gamma-ray band and even giant flares, which could be highly super-Eddington. 
Besides, AXPs/SGRs have long spin periods clustered in the range of 2 -- 12 s, and their period derivatives are large too.

Conventional models for AXPs/SGRs are magnetars~\citep{Duncan92}, isolated neutron stars with extremely strong dipole and multipole magnetic fields (higher than the quantum critical magnetic field $B_{\rm QED} = m^2_ec^3/e\hbar = 4.4 \times 10^{13}$ G). 
The persistent emission is powered by magnetic field decay; magnetic dipole radiation would contribute to their spin-down, and the effects of twisted magnetosphere~\citep{Thompson02} and wind braking~\citep{Tong13} have also been considered.
%
Sudden release of magnetic energy, such as magnetic reconnection, could result in bursts or giant flares.
Although magnetar model could explain some of the properties of AXPs/SGRs, it is still facing some problems arising from accumulating observations: few predictions have been confirmed yet.
%
%
%
%
%
%
After all, there is no direct evidence for the existence of super-strong magnetic field.
Alternative models of AXPs/SGRs are not only possible but also welcome.

AXPs/SGRs are suggested to be normal-field pulsar-like stars accreting from supernova fallback disks~\citep{Chatterjee00, Alpar01}.
Accretion energy could power the persistent emission, and propeller effect may account for the braking mechanism, as well as the period clustering of AXPs/SGRs.
However, fallback disk models could not explain the super-Eddington bursts or giant flares.
The problem could be solved if the compact star is a solid quark star~\citep{Xu03},  since the self-confined surface~\citep{Alock86} could explain the super-Eddington phenomena, and energy released during star quakes~\citep{Xu06} may be alternative power source for bursts and giant flares.
Therefore, AXPs/SGRs could be quark star/fallback disk systems~\citep{Xu07, TX11}.

%
%
To determine whether AXPs/SGRs are magnetars or fallback disk systems is of fundamental importance.
It could help us understand the observational phenomena of AXPs/SGRs, and even give hints on the nature of pulsar-like stars, which is related to the state of cold matter at supra-nuclear density and strong interaction.
In this paper, we would like to study the problem from the point of X-ray spectrum of AXPs/SGRs.

AXPs/SGRs have soft spectra below 10 keV that are generally fitted by a combination of a steep power-law with photon index  $\Gamma \sim 2-4$ and a blackbody with temperature $kT \sim 0.5$ keV~\citep{Mereghetti08}.
Non-thermal hard X-ray components above 15 keV in AXPs/SGRs were discovered in recent years, with different spectral properties from the soft X-ray band~\citep{Kuiper06}.
The hard X-ray spectra are well fitted by flat power-laws with photon index $\Gamma \sim 0.5-1.5$, and the luminosity is similar to that of the soft X-ray band.
Therefore, the hard X-rays provide us important information to understand the magnetic fields and surface properties, and could put strong constraints on the theoretical modeling of AXPs/SGRs.
The physical mechanism of hard X-ray emission is still unknown, but some possibilities have been proposed trying to explain it.

In the frame of magnetars, quantum electrodynamics model~\citep{Heyl05}, bremsstrahlung model~\citep{Beloborodov07} and resonant inverse Compton scattering model~\citep{Baring07} have been explored, predicting power-law spectra with different cutoff properties.
%
%
%
%
The spectral cutoff properties of AXPs/SGRs are not well understood yet.
Upper limits in MeV bands are obtained with the archival CGRO COMPTEL data of four AXPs, which indicate cutoff energy below 1 MeV~\citep{Kuiper06}.
%
%
Accumulating INTEGRAL IBIS data over 9 years, the time-averaged spectrum of 4U 0142+61, the brightest AXP,  can be fitted with a power-law with an exponential cutoff at $\sim 130$ keV, which might rule out models involving ultra-relativistic electrons~\citep{Wang13}.
In the fallback disk frame, ~\citet{Trumper10} considered producing the hard X-ray emission by bulk-motion Comptonization (BMC) process of surface photons in the accretion flow.
Their work shows that for 4U 0142+61, BMC model could reproduce both the soft and hard X-ray spectra.
BMC model is successful in explaining the spectra of 4U 0142+61, but its applicability to other sources remains in doubt.
%
%

In this work, we try to apply BMC model to other AXPs/SGRs, and make simulation to discuss how to distinguish different models by future observations.
Using data from {\it Suzaku} and INTEGRAL, we derive the soft and hard X-ray spectra of four sources, namely AXP 1RXS J170849-400910, AXP 1E 1547.0-5408, SGR 1806-20 and SGR 0501+4516 (hereafter abbreviated as 1RXS J1708-40, 1E 1547-54, SGR 1806-20 and SGR 0501+45), to put further constraints on BMC model.
We find that the spectra of all the chosen sources can be well fitted with XSPEC model compTB, showing that accretion scenario could be compatible with X-ray emission of AXPs/SGRs.
%
To investigate the feasibility of discriminating various models of hard X-ray emission by future observations, we also make simulation for the hard X-ray modulation telescope (HXMT) \footnote{http://heat.tsinghua.edu.cn/hxmtsci/hxmt.html}.
Simulated spectra of BMC model exhibit cutoff around 200 keV, which could distinguish BMC from other cases in the magnetar model.
%

In \S~2 we will introduce the utilized {\it Suzaku} and INTEGRAL observations, along with data analysis including spectral properties and time variabilities.
Then we present the averaged spectra and the fitting of compTB model in \S~3. 
HXMT simulations are shown in \S~4, and possible discrepancies between various models of the hard tail are also discussed.
Finally we make conclusions in \S~5. 

\section{observations and data analysis}

\subsection{Source selection}

{\it Suzaku} provides for the first time simultaneously observed soft and hard X-ray spectra of seven AXPs/SGRs~\citep{Enoto10}.
%
Among the seven sources, SGR 1900+14 is only detected up to $\sim$ 50 keV due to its relatively low flux, and the soft X-ray spectrum of 1E 1841-045 is contaminated by emission lines from surrounded supernova remenants~\citep[see][Fig. 1]{Enoto10}. 
Therefore, in our analysis we will only focus on four sources, namely 1RXS J1708-40, 1E 1547-54, SGR 1806-20 and SGR 0501+45, while the data of 4U 0142+61, SGR 1900+14 and 1E 1841-045 are not analyzed.
%
%
%
For the hard X-ray spectra, INTEGRAL observations are also used, which could reach higher energy and place better constraints on parameters.
Because of the lower sensitivity of INTEGRAL hard X-ray detector than that of {\it Suzaku}, spectrum of single observation has to be summed up to get an acceptable signal-to-noise ratio (S/N).
The spectral fitting software used is XSPEC version 12.8.0, and all cited errors are at $1 \sigma$ level.

\subsection{{\it Suzaku} observations}

{\it Suzaku} observations utilized for the four sources are listed in Table~\ref{obs}.
We extracted spectra using data from the X-ray Imaging Spectrometer (XIS; \citealt{XIS}) and the Hard X-ray Detector (HXD; \citealt{HXD}), which are sensitive in the energy range of 0.2--12 keV and 10--70 keV.
The data reduction was carried out using HEASOFT version 6.13.
The XIS and HXD data were reprocessed with the pipeline processing version 2.4, employing recommended data screening criteria. 
We accumulated screened data of XIS from a region within 2' radius of the source centroid, and derived background spectrum from source-free regions in the immediate vicinity of the target.
For HXD-PIN data, non-X-ray background (NXB) and cosmic X-ray background (CXB) were subtracted to obtain spectra.%

\begin{table}
\bc
\begin{minipage}[]{100mm}
\caption[]{{\it Suzaku} and INTEGRAL Observations. 
\label{obs}}\end{minipage}
\setlength{\tabcolsep}{5pt}
 \begin{tabular}{ccccccc}
  \hline \hline \noalign{\smallskip}
  & {\it Suzaku} & & & INTEGRAL \\
Name &  ObsID & Time span & Exposure & Revs. & Time span & Exposure  \\
          & & (yy/mm/dd) &  (ks) & & (yy/mm/dd) & (Ms)\\
  \hline\noalign{\smallskip}
1RXS J1708-40 & 404080010 & 09/08/23 -- 09/08/24 & 47.9 & 0037 -- 1088 & 03/02/01 -- 11/09/13 & 5.84\\
1E 1547-54 & 903006010 & 09/01/28 -- 09/01/29 & 31.0 & 0767 -- 0769 & 09/01/28 -- 09/02/01 & 0.16 \\
SGR 1806-20 & 402094010 & 07/10/14 -- 07/10/15 & 46.6 & 0286 -- 1080 & 05/02/16 -- 11/08/20 & 4.24 \\ 
SGR 0501+45 & 903002010 & 08/08/26 -- 08/08/27 & 50.7 & 0047 -- 1141 & 03/03/03 -- 12/02/17 & 1.45  \\
  \noalign{\smallskip}\hline
\end{tabular}
\ec
\end{table}

We fit the soft X-ray spectra with a two component blackbody plus power-law model affected by photoelectric absorption, and the hard X-ray spectra with a single power-law model.
The fitting results, as shown in Table~\ref{fit_bp}, of 1RXS J1708-40 and SGR 1806-20 are in agreement with values from other observations of the sources within error~\citep{Rea07, Esposito07,Enoto10}. 
Since 1E 1547-54 and SGR 0501+45 are in outburst during the {\it Suzaku} observations used, it is not feasible to compare the fitting parameters with observations in quiescent states.
So we fit the broad band spectra of 1E 1547-54 with a two component blackbody plus power-law model, while a three component model including two blackbodies and a power-law is applied to SGR 0501+45, as tried in previous analysis of the same observations~\citep{suzaku_1547,suzaku_0501}, and the parameters are also consistent with former results.

\begin{table}
\bc
\begin{minipage}[]{100mm}
\caption[]{{Soft and hard X-ray spectral properties of {\it Suzaku} data. } 
\label{fit_bp}}
\setlength{\tabcolsep}{3pt}
\end{minipage}
 \begin{tabular}{ccccccc}
  \hline \hline \noalign{\smallskip}
Name & Model$^{\rm a}$ &  $N_{\rm H}$ & $kT$ & $\Gamma$ & $\chi^2/d.o.f.$ & Flux$^{\rm b}$\\
 & & ($10^{22}$ cm$^{-2}$) & (keV) \\
  \hline\noalign{\smallskip}  
1RXS J1708-40 & BB+PL & $1.40 \pm 0.04$ & $0.45\pm 0.01$ & $2.72 \pm 0.07$ & 1.23 (169) & $3.17 \pm 0.01$  \\
1E 1547-54 & BB+PL & $3.46_{-0.22}^{+0.21}$ & $0.61_{-0.03}^{+0.04}$ & $2.38_{-0.22}^{+0.19}$ & 1.01 (87) & $6.63 \pm 0.07$ \\
SGR 1806-20 & BB+PL & $8.43_{-0.96}^{+1.07}$ & $0.40^{+0.09}_{-0.06}$ & $1.97_{-0.20}^{+0.17}$ & 1.09 (80) & $0.95 \pm 0.02$ \\
SGR 0501+45 & BB+PL & $1.06 \pm 0.02$ & $0.69 \pm 0.01$ & $2.96 \pm 0.04$ & 1.11 (231) & $2.94 \pm 0.01$ \\
\hline \noalign{\smallskip}
1RXS J1708-40 & PL & -- & -- & $1.73 \pm 0.18$ & 0.94 (9) & $2.59 \pm 0.33$ \\
1E 1547-54 & PL & -- & -- & $1.55 \pm 0.08$ & 1.05 (16) & $9.34 \pm 0.05$\\
SGR 1806-20 & PL & -- & -- & $1.69 \pm 0.15$ & 0.94 (16) & $2.98 \pm 0.32$ \\
SGR 0501+45 & PL & -- & -- & $0.43 \pm 0.26$ & 0.99 (7) & $3.05 \pm 0.45$\\
\hline \noalign{\smallskip}
1E 1547-54 & BB+PL & $2.81 \pm 0.08$ & $0.65 \pm 0.02$ & $1.54 \pm 0.05$ & 1.10 (105) & $6.83 \pm 0.05/9.39 \pm$ 0.04 \\
SGR 0501+45 & 2BB+PL & $0.73 \pm 0.03$ & $0.33 \pm 0.02/0.73\pm 0.01$ & $1.35_{-0.11}^{+0.12}$ & 1.13 (238) & $2.95 \pm 0.01/1.90 \pm 0.19$ \\

  \noalign{\smallskip}\hline
\end{tabular}
\ec
\tablecomments{0.86\textwidth}{The first four rows are BB+PL fitting to the soft X-ray spectra; the middle four rows are PL fitting to the hard X-ray spectra; the last two rows are fitting results of broad band X-ray spectra. \\
$^{\rm a}$ BB and PL represent blackbody and power-law respectively.\\
$^ {\rm b}$ 2--10 keV flux for soft spectra and 20--60 keV flux for hard spectra, in units of $10^{-11}$ erg cm$^{-2}$ s$^{-1}$.}

\end{table}

\subsection{INTEGRAL observations}

The gamma-ray mission INTErnational Gamma-Ray Astrophysics Laboratory (INTEGRAL; ~\citealt{INTEGRAL}) has been operational since October 2002.
Its imager IBIS (Imager on Board the INTEGRAL Satellite; ~\citealt{IBIS}) has a low-energy detector sensitive between 20--300 keV, the INTEGRAL Soft Gamma-Ray Imager (ISGRI; ~\citealt{ISGRI}).
We used all public data from the INTEGRAL Science Data Center (ISDC) where the sources are within 10 degrees of the pointing direction from 2003 to 2011. 
The IBIS-ISGRI data were reduced with the Off-line Scientific Analysis (OSA) software version 10.
Following the standard procedures described in the IBIS analysis user manual, we created images in four energy bands (20--40 keV, 40--60 keV, 80--100 keV, 100--300 keV) for source detection, and extracted spectra of the sources for each individual pointing.
%
%

\begin{figure}
   \centering
     \includegraphics[width=0.42\textwidth]{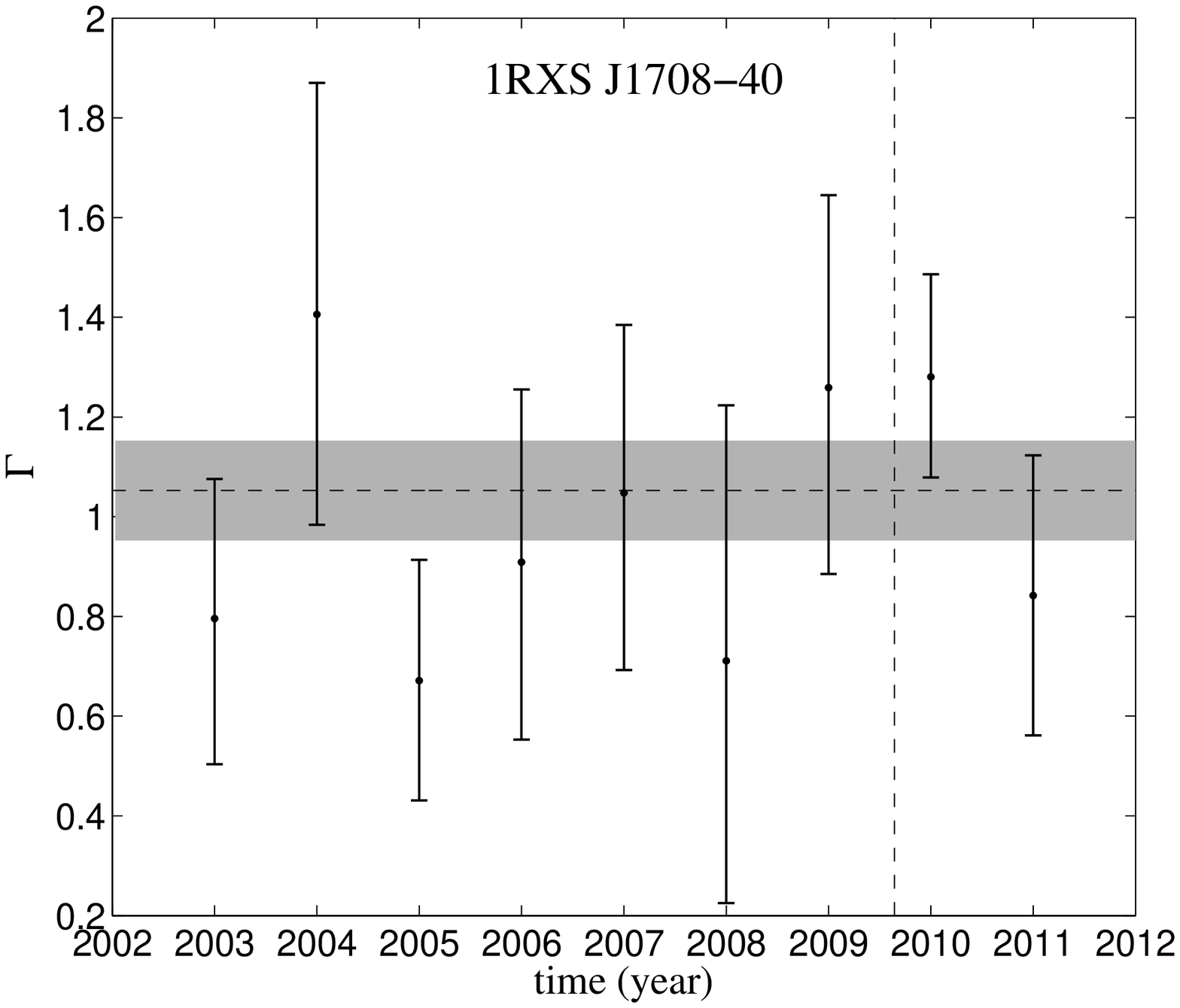}
     \includegraphics[width=0.42\textwidth]{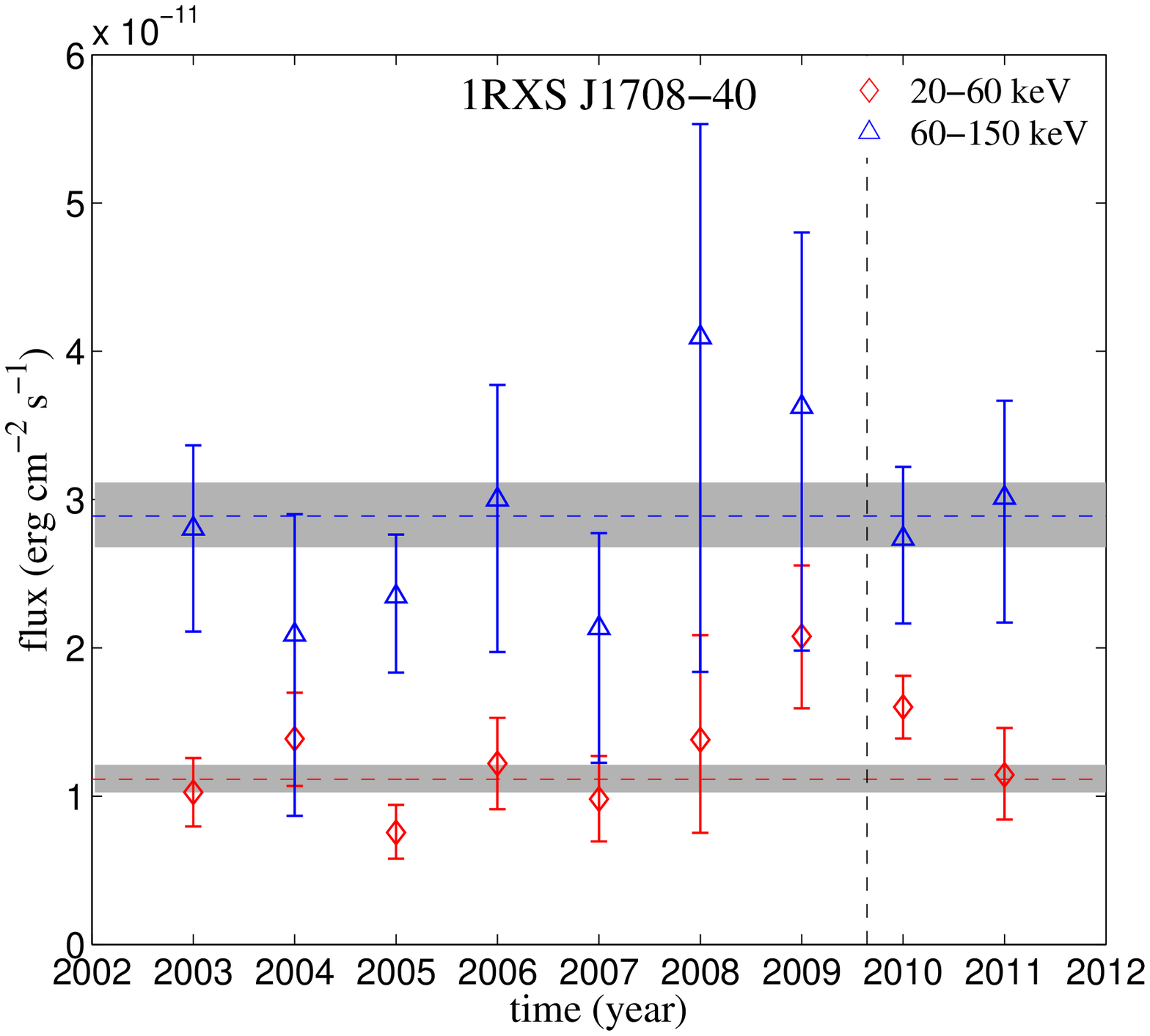}
     \includegraphics[width=0.42\textwidth]{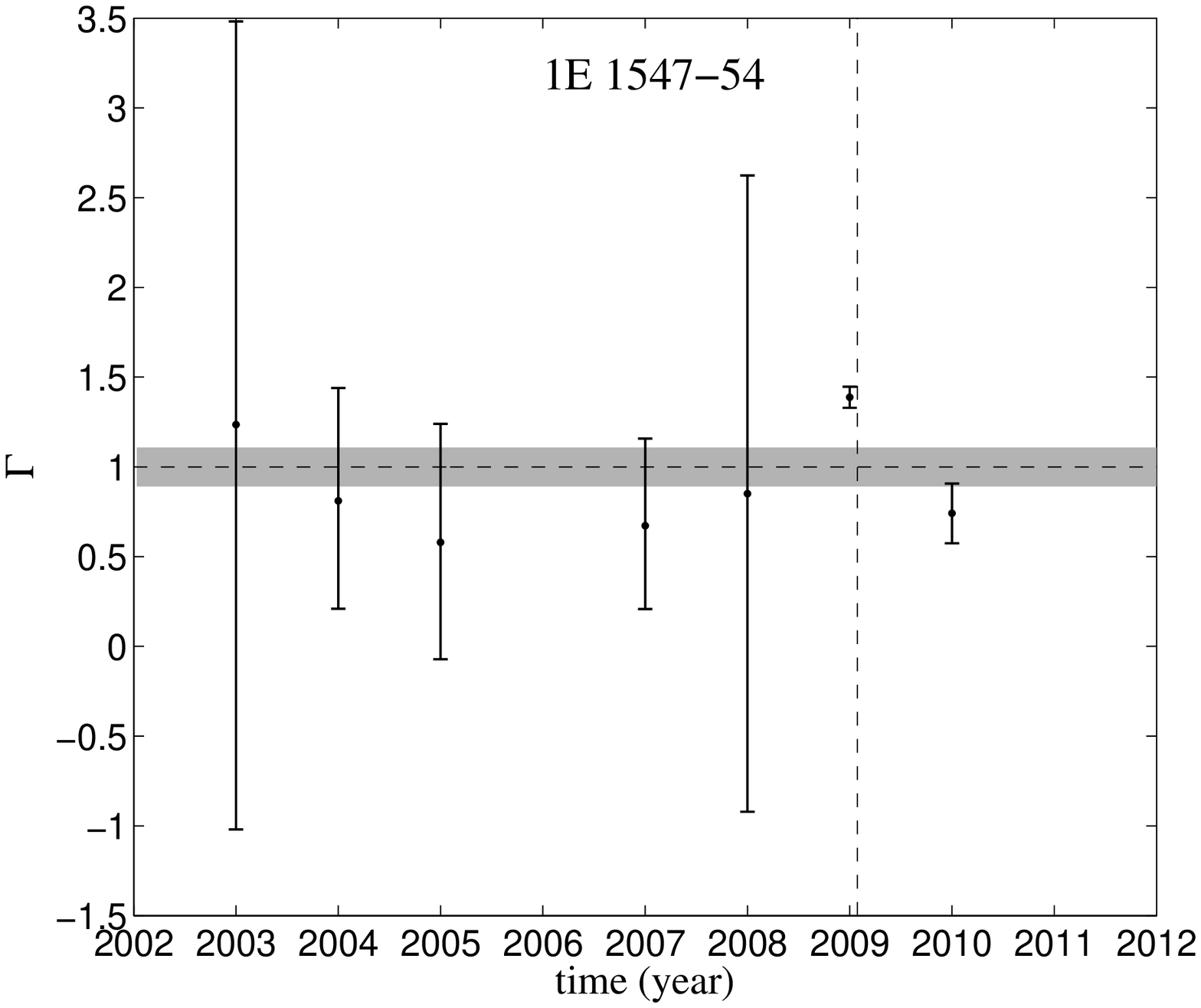}
     \includegraphics[width=0.42\textwidth]{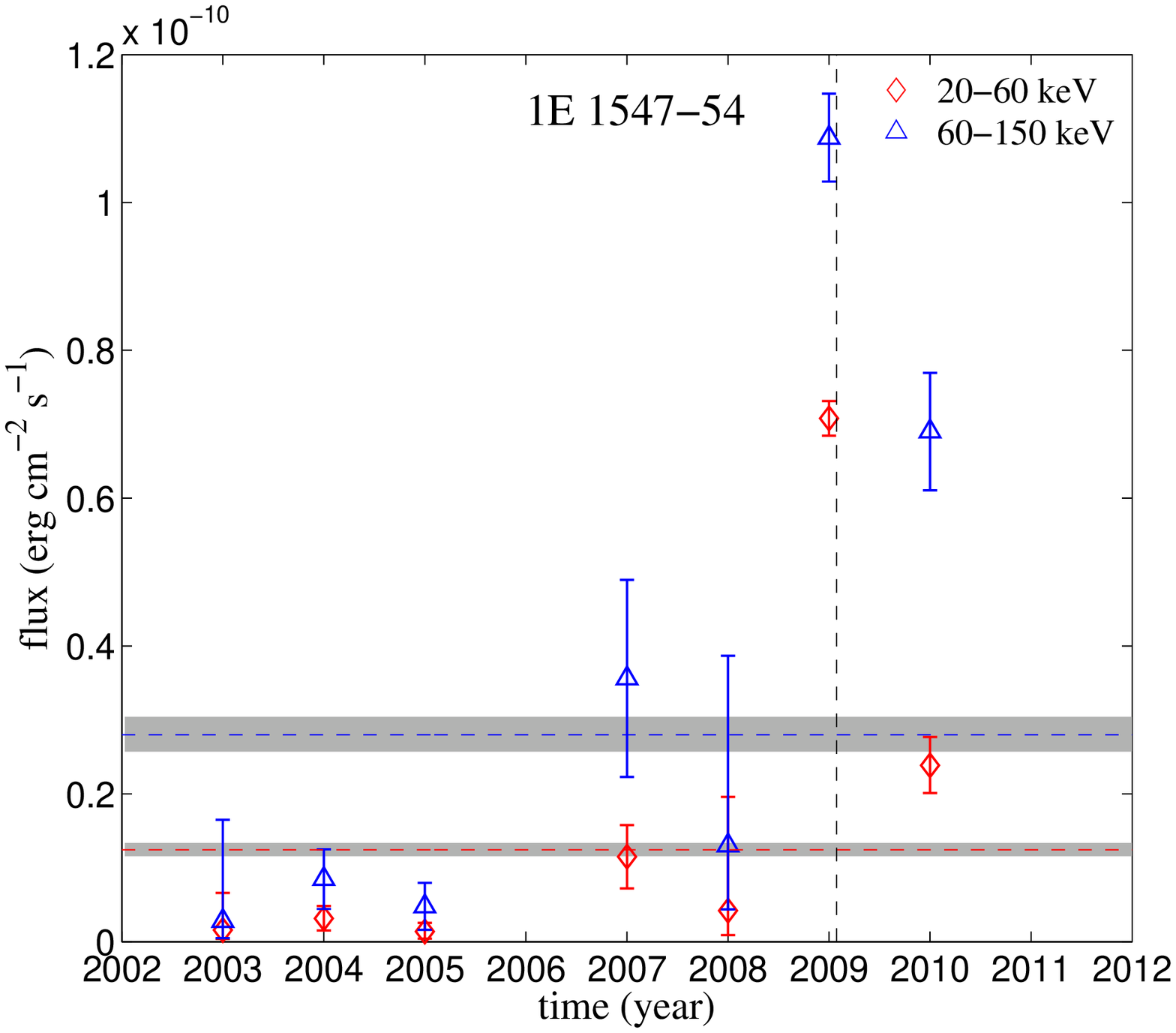}
     \includegraphics[width=0.42\textwidth]{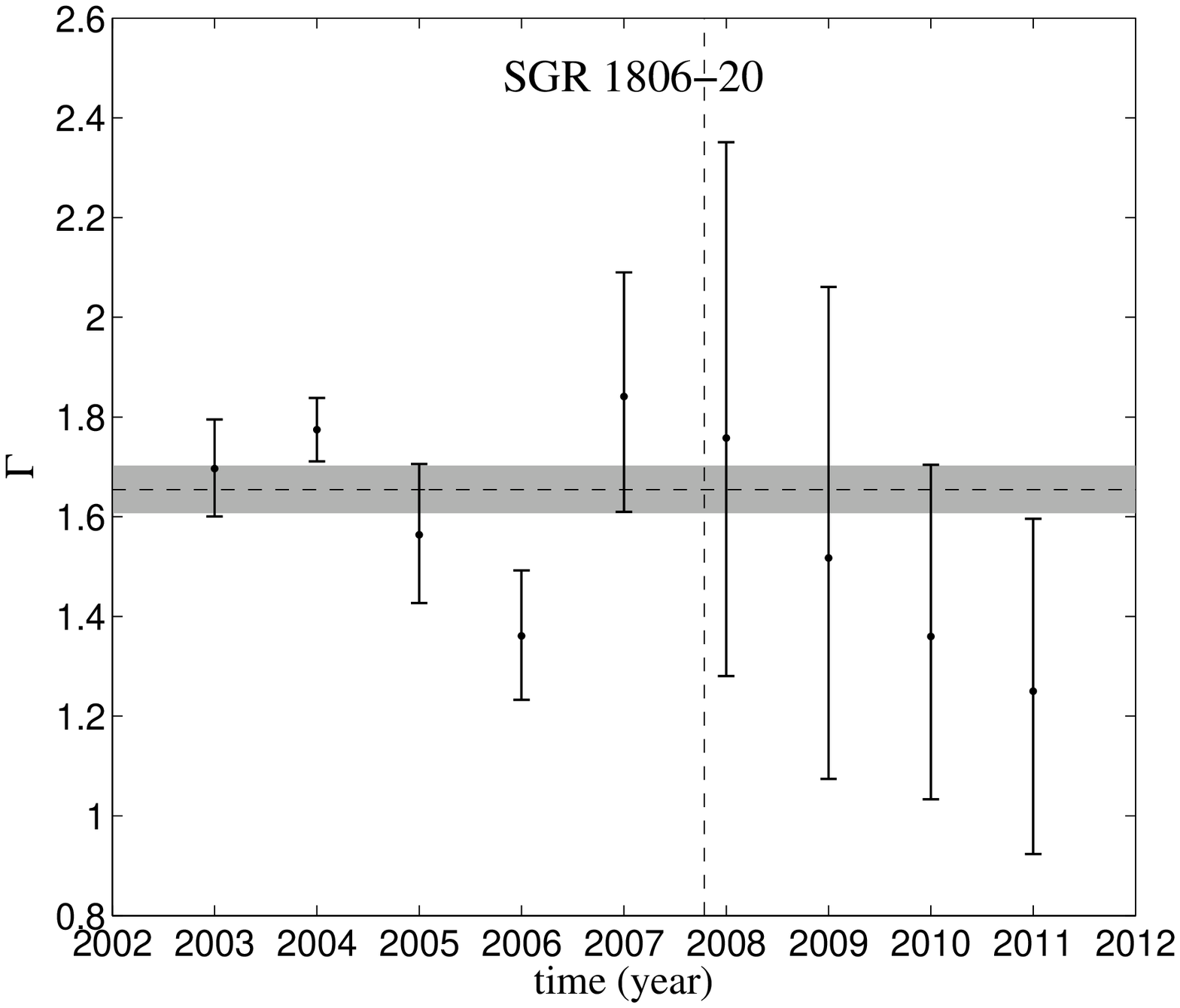}
     \includegraphics[width=0.42\textwidth]{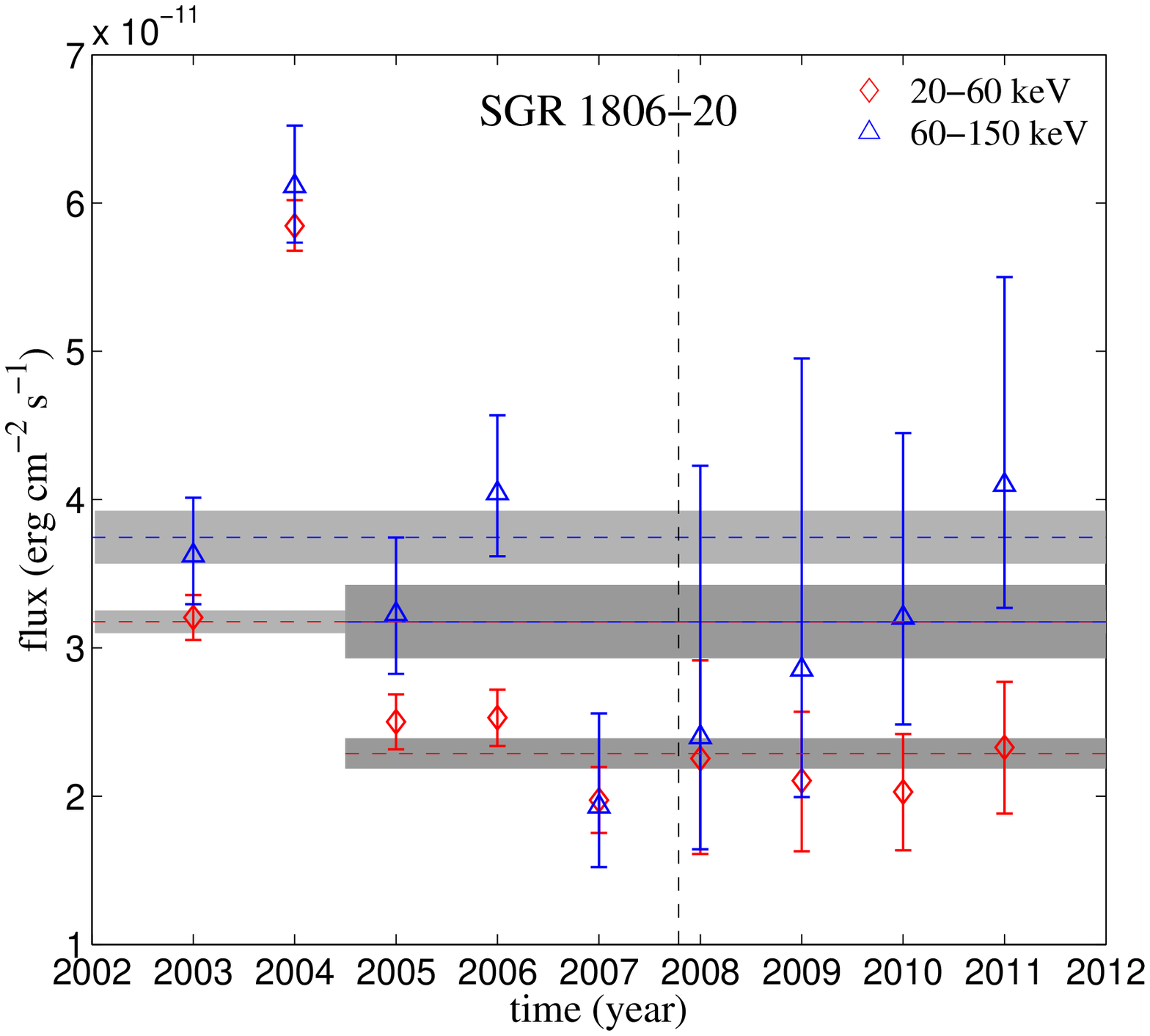}

   \caption{Photon indices (left panels) and fluxes (right panels) for every one-year time interval of 1RXS J1708-40, 1E 1547-54 and SGR 1806-20. Horizontal dotted lines represent time-averaged values, with their 1$\sigma$ errors indicated by grey bands, and vertical dotted lines indicate the observation time of {\it Suzaku}. Fluxes in the energy range of 20--60 keV and 60--150 keV are in red (with diamond markers) and blue (with triangle markers) respectively. For SGR 1806-20, time-averaged values from 2005 to 2011 are also plotted.}
   \label{fvar}
   \end{figure}

To get a better S/N, spectra of individual pointings are summed up to derive the time-averaged spectra.
Even so, the time-averaged spectrum of SGR 0501+45 is of poor quality (see bottom-right panel in Figure~\ref{bmc_i}), so we only analyze INTEGRAL spectra of the other three sources.
Considering possible variability of the sources in such a long time, we divide the INTEGRAL observations from 2003 to 2011 into one-year time intervals and extract spectra respectively. 
%
The spectra are fitted with single power-law model, and the fitting results as well as 20--150 keV fluxes are listed in Table~\ref{tvar}.
We also plot the photon indices $\Gamma$ and fluxes in Figure~\ref{fvar}.

For 1RXS J1708-40, 
%
the deviation of photon indices and fluxes from the time averaged value is within $2\sigma$ level, thus there is no significant variation.
In case of SGR 1806-20, the fluxes in 2003 and 2004 are apparently higher than those of the other years, presumably due to its giant flare in 2004~\citep{Hurley05}.
In the following seven years, the fluxes and photon indices do not change significantly.
%
So we only sum up spectra of observations from 2005 to 2011.

However, the fluxes of 1E 1547-54 vary dramatically.
From 2003 to 2008, the source is not bright in hard X-ray band, 
and an outburst is observed in 2009, during which its hard X-ray emission was first discovered by {\it Suzaku}. 
For compatibility with Suzaku soft X-ray data of 1E 1547-54, the INTEGRAL observations utilized are only Revs. 768-769, from Jan. 28, 2009 
to Feb. 1, 2009 
, which overlap in time with {\it Suzaku} observations from Jan. 28, 2009 
to Jan. 29, 2009. 
As both the soft and hard spectra vary slightly from Jan. 28, 2009 to Feb. 7, 2009~\citep{Bernardini11}, the longer time span of INTEGRAL observations than that of {\it Suzaku} would make little difference.

%
The selection of INTEGRAL data for the three sources to get averaged spectra is analyzed above, and we summarize the resulting total exposure time in Table~\ref{obs}.
We also fit the averaged spectra with power-law model, and the fitting results are given in Table~\ref{tvar}.

\begin{table}
\bc
\begin{minipage}[]{100mm}
\caption[]{{Time variations of INTEGRAL observations.} \label{tvar}}
\end{minipage}
\setlength{\tabcolsep}{3pt}
 \begin{tabular}{cccccccccc}
  \hline \hline \noalign{\smallskip}
 & 1RXS J1708-40 & & & 1E 1547-54 & & & SGR 1806-20 & & \\
Year & $\Gamma$ & Flux$^{\rm a}$ & $\chi^2/d.o.f.$ & $\Gamma$ & Flux$^{\rm a}$ & $\chi^2/d.o.f.$ & $\Gamma$ & Flux$^{\rm a}$ & $\chi^2/d.o.f.$  \\
  \hline\noalign{\smallskip}  
2003 &  $0.79_{-0.29}^{+0.28}$ & $3.82 \pm 0.64$  & 1.25 (9) & $1.24 \pm 2.26$  & $0.43_{-0.35}^{+1.74}$  & 1.71 (9) & $1.70 \pm 0.10$ & $6.83_{-0.22}^{+0.23}$  & 0.87 (9)  \\
2004 & $1.40_{-0.42}^{+0.46}$ & $3.48_{-0.88}^{+0.89}$ & 0.60 (9) & $0.81_{-0.63}^{+0.60}$ & $1.16 \pm 0.49$ & 0.91 (9) & $1.77 \pm 0.06$ & $12.0 \pm 0.2$ & 1.72 (9) \\
2005& $0.67 \pm 0.24$ &  $3.10 \pm 0.52$ & 1.10 (9) & $0.58 \pm 1.09$ & $0.61_{-0.66}^{+0.65}$ & 1.64 (9) & $1.56 \pm 0.14$ & $5.73_{-0.30}^{+0.31}$ & 0.74 (9) \\
2006 & $0.91_{-0.36}^{+0.34}$ & $4.22 \pm 0.87$ & 1.21 (9) & -- & -- & -- & $1.36 \pm 0.13$ & $6.57 \pm 0.33$ & 0.87 (9)\\
2007 & $1.05_{-0.36}^{+0.34}$ & $3.10_{-0.79}^{+0.80}$ & 1.10 (9) & $0.67_{-0.49}^{+0.46}$ & $4.71 \pm 1.76$ & 1.25 (9) & $1.84_{-0.23}^{+0.25}$ & $3.90_{-0.30}^{+0.31}$ & 0.58 (9) \\
2008 & $0.71_{-0.49}^{+0.51}$ & $5.47 \pm 1.83$ & 0.70 (9) & $0.85 \pm 1.77$ & $1.73_{-1.10}^{+3.01}$ & 1.96 (9) & $1.78_{-0.48}^{+0.59}$ & $4.65_{-0.67}^{0.68}$ & 1.69 (9) \\
2009 & $1.26_{-0.37}^{+0.39}$ & $5.70_{-1.31}^{+1.33}$ & 1.13 (9) & $1.39 \pm 0.06$ & $18.0 \pm 0.7$ & 0.83 (9) & $1.52_{-0.44}^{+0.54}$ & $4.95_{-0.79}^{+0.78}$ & 1.09 (9) \\
2010 & $1.28_{-0.20}^{+0.21}$ & $4.34 \pm 0.56$ & 1.23 (9) & $0.74 \pm 0.17$ & $9.29 \pm 0.99$ & 0.75 (9) & $1.36_{-0.33}^{+0.34}$ & $5.23_{-0.63}^{+0.64}$ & 0.92 (9) \\
2011 & $0.84 \pm 0.28$ & $4.16_{-0.84}^{+0.83}$ & 1.49 (9) & -- & -- & -- & $1.25_{-0.33}^{+0.35}$ & $6.42 \pm 0.71$ & 1.15 (9) \\ 
average & $0.95 \pm 0.12$ & $4.01 \pm 0.23 $ & 1.82 (6) & $1.57 \pm 0.09$ & $21.0 \pm 1.3$ & 1.07 (8) &  $1.49 \pm 0.08$ & $5.46 \pm 0.29$ & 1.44 (8) \\
  \noalign{\smallskip}\hline
\end{tabular}
\ec
\tablecomments{0.86\textwidth}{$^ {\rm a}$ 20--150 keV flux in units of $10^{-11}$ erg cm$^{-2}$ s$^{-1}$.}
\end{table}



\section{application of BMC model to the averaged spectra}

If AXPs/SGRs are fallback disk systems, soft photons emitted from polar cap would be in the accretion flow.
Some of the seed photons get upscattered by Comptonization process with high-energy electrons, producing the hard X-ray emission, while others escape directly and constitute the soft X-ray spectra~\citep{Trumper10}.
In spectral fitting, the BMC process is described by the XSPEC model compTB~\citep{Farinelli08}, in which thermal and bulk-motion Comptonization of seed photons are considered. 
%
This model consists of two components, the direct seed photon spectrum, and the Comptonizated spectrum obtained as a self-consistent convolution of the seed spectrum with the system's Green function.
The seed photon spectrum is a modified blackbody $S(E) \propto E^\gamma / (e^{E/kT_s}-1)$, where $kT_s$ is the blackbody temperature and $\gamma$ represents a modification of blackbody.
The Comptonization process is characterized by three parameters, bulk parameter $\delta=<E_{bulk}>/<E_{th}>$ describing the relative efficiency of bulk over thermal Comptonization, electron temperature $kT_e$ and energy index of the Comptonization spectrum $\alpha$. 
There are also two coefficients, illumination factor $A$ and normalization of the seed photon spectrum $C_N$.
%


First we fit the broad band {\it Suzaku} spectra of the four sources with compTB, which provides good fits from the statistical point of view.
However, when initial values of parameters are changed, the spectra fitting could give different result with similar $\chi^2$.
We list some fitting results in Table~\ref{fit_s} in the order of increasing $\delta$. 
The parameters are poorly constrained, with problem mainly lying in the  degeneracy of $\delta$ and $kT_e$.
$\delta$ ranges from $\sim 1$ to $\sim 100$, and $kT_e$ ranges from $\sim 1$ keV to $\sim 10$ keV.
This should be attributed to the low S/N of {\it Suzaku} data in hard X-ray band, while the equivalence of BMC and TC process in upscattering seed photons might cause such situation to some extent.

The parameter $\gamma$ is left free in spectral fitting, and its best-fit values are in the range of  0.18 -- 1.2, indicating a significant modification of the seed spectrum. 
Such large deviation of seed spectrum from pure blackbody is questionable, so we freeze $\gamma$ at 3 and try to fit the spectra with compTB again.
The spectra of 1E 1547-54 and SGR 1806-20 can be fitted with a larger $\chi^2$, and $kT_s$ is similar to the temperature of blackbody plus power-law fit to soft X-ray data.
However, 1RXS J1708-40 and SGR 0501+45 can not get acceptable fits.
On the other hand, the broad band spectra of 1E 1547-54 and SGR 1806-20 could be fitted with an absorbed blackbody plus power-law model, while an additional blackbody component is required for the spectral fitting of 1RXS J1708-40 and SGR 0501+45.

\begin{table}
\bc
\begin{minipage}[]{100mm}
\caption[]{The fitting results of {\it Suzaku}  data with compTB model.   
\label{fit_s}}
\end{minipage}
 \begin{tabular}{cccccc}
  \hline\noalign{\smallskip}
 & 1RXS J1708-40 & & 1E 1547-54 & &  \\
  \hline\noalign{\smallskip}
$N_{\rm H} (10^{22} {\rm cm}^{-2})$ & $1.50 \pm 0.01 $ & $1.50 \pm 0.01$ & $3.42 \pm 0.03$ & $3.40 \pm 0.03$ & $3.26 \pm 0.03$\\
$kT_{\rm s}$ (keV) & $1.08 \pm 0.01$ & $1.07 \pm 0.01$ & $1.17 \pm 0.02$ & $1.12 \pm 0.02$ & $1.00 \pm 0.02$  \\
$\gamma$ & $0.22_{-0.14}^{+0.12}$ & $0.26_{-0.17}^{+0.01}$ & $0.79 \pm 0.04$  & $0.86 \pm 0.01$ & $1.20 \pm 0.02$ \\
$\alpha$ & $0.60 \pm 0.01$ & $0.62 \pm 0.01$  & $0.60 \pm 0.01$ & $0.67 \pm 0.01$ & $0.68 \pm 0.01$ \\ 
$\delta$ & $19.1_{-3.9}^{+8.5}$ & $44.0_{-7.0}^{+15.4}$ & $1.34_{-1.21}^{+3.04}$ & $79.3_{-7.3}^{+10.2}$ & $145_{-13}^{+18}$ \\
$kT_e$ (keV) & $3.87 \pm 0.14$ & $1.57 \pm 0.06$ & $31.1_{-4.6}^{+5.0}$ & $0.89_{-0.09}^{+0.13}$ & $0.47_{-0.04}^{+0.06}$ \\
log(A) & $-0.91 \pm 0.01$ & $-0.58 \pm 0.01$ & $-0.47 \pm 0.01$ & $-0.14 \pm 0.01$ & $-0.41 \pm 0.01$ \\
$\chi_r^2$ (d.o.f.) & 1.109 (177) & 1.108 (177)  & 1.003 (102) & 0.982 (102) & 0.983 (102) \\
  \noalign{\smallskip}\hline
  
  \hline\noalign{\smallskip}
 & SGR 1806-20 & SGR 0501+45 & &  \\
  \hline\noalign{\smallskip}
$N_{\rm H} (10^{22} {\rm cm}^{-2})$ & $7.80_{-0.14}^{+0.15}$ & $0.80 \pm 0.01$ & $0.79 \pm 0.01 $ & $0.79 \pm 0.01 $ \\
$kT_{\rm s}$ (keV) & $1.06 \pm 0.02$ & $1.11 \pm 0.01$ & $1.09 \pm 0.01$ & $1.08 \pm 0.01$  \\
$\gamma$ & $0.20 \pm 0.01$ & $1.23 \pm 0.01$ & $1.27 \pm 0.01$ & $1.28 \pm 0.01$  \\
$\alpha$ & $0.32 \pm 0.01$ & $0.07 \pm 0.01$ & $0.28_{-0.03}^{+0.04}$ & $0.30_{-0.03}^{+0.04}$  \\ 
$\delta$ & $200 (>9.23)$ & $1.88_{-0.41}^{+0.70}$ & $18.6_{-1.9}^{+2.8}$ & $189 (>10.8)$  \\
$kT_{\rm e}$ (keV) & $34.2_{-5.8}^{+7.2}$ & $43.2_{-5.5}^{+10.3}$ & $3.29_{-0.31}^{+0.46}$ & $0.33_{-0.02}^{+0.04}$  \\
log(A) & $-1.44 \pm 0.02$ & $-0.99 \pm 0.02$ & $-1.25 \pm 0.02$ & $-1.24 \pm 0.02$  \\
$\chi_r^2$ (d.o.f.) & 1.008 (95) & 1.14 (237)  & 1.13 (237) & 1.13 (237)  \\
  \noalign{\smallskip}\hline

\end{tabular}
\ec
\end{table}

%
%
%

\begin{figure}
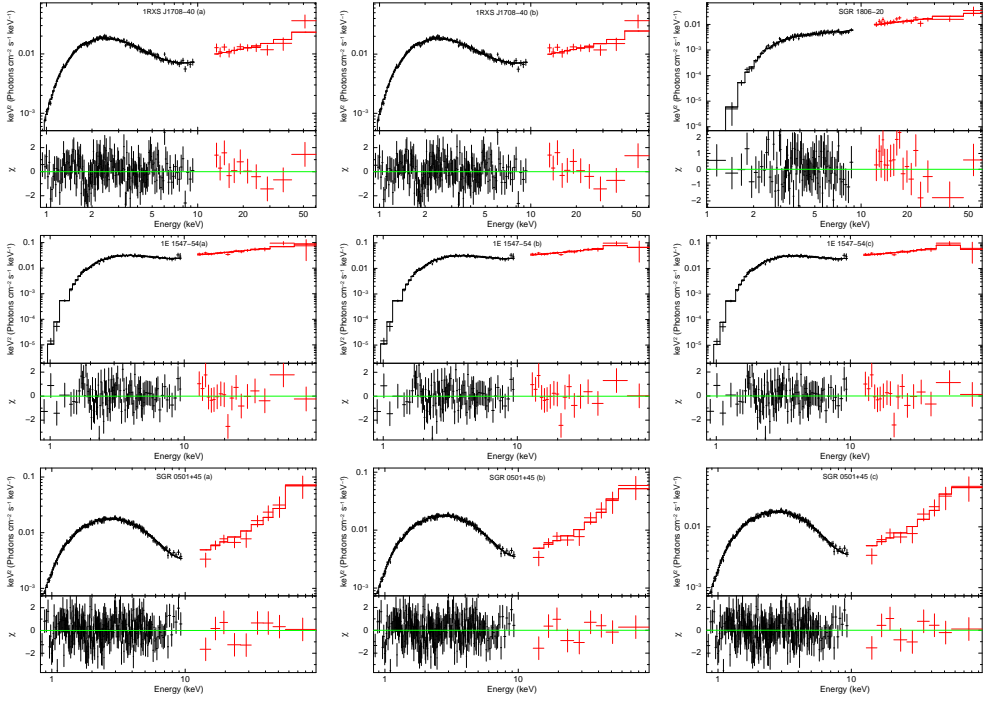

     \includegraphics[height=0.3\textwidth, angle=-90]{1708bmcs_a.eps}
     \includegraphics[height=0.3\textwidth, angle=-90]{1708bmcs_b.eps}
     \includegraphics[height=0.3\textwidth, angle=-90]{1806bmcs.eps}

     \includegraphics[height=0.3\textwidth, angle=-90]{1547bmcs_a.eps}
     \includegraphics[height=0.3\textwidth, angle=-90]{1547bmcs_b.eps}
     \includegraphics[height=0.3\textwidth, angle=-90]{1547bmcs_c.eps}
     
     \includegraphics[height=0.3\textwidth, angle=-90]{0501bmcs_a.eps}
     \includegraphics[height=0.3\textwidth, angle=-90]{0501bmcs_b.eps}
     \includegraphics[height=0.3\textwidth, angle=-90]{0501bmcs_c.eps}

   \caption{$E^2f(E)$ spectra of the four studied AXPs/SGRs, along with the best-fit compTB models and residuals in units of $\sigma$. Suzaku XIS and HXD data are in black and red respectively.}
   \label{bmc_s}
   \end{figure}

\begin{figure}
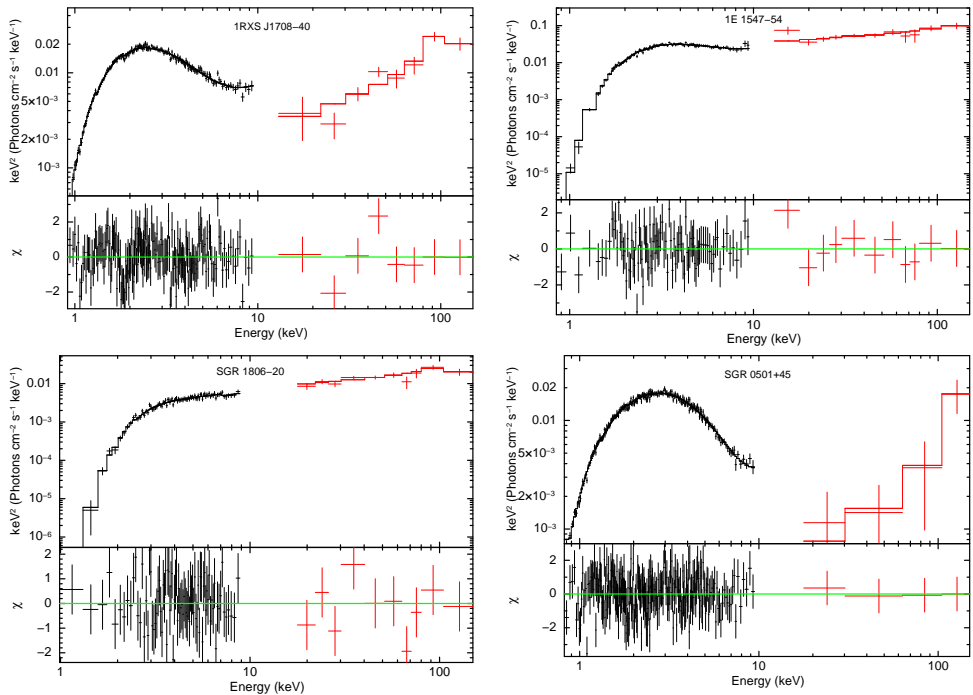

     \includegraphics[height=0.45\textwidth, angle=-90]{1708spe.eps}
     \includegraphics[height=0.45\textwidth, angle=-90]{1547spe.eps}\\
     \includegraphics[height=0.45\textwidth, angle=-90]{1806spe.eps}
     \includegraphics[height=0.45\textwidth, angle=-90]{0501spe.eps} \\

   \caption{$E^2f(E)$ spectra of the four studied AXPs/SGRs, along with the best-fit compTB models and residuals in units of $\sigma$. Suzaku XIS and INTEGRAL IBIS-ISGRI data are in black and red respectively.}
   \label{bmc_i}
   \end{figure}

To better constrain the parameters, we replace the hard X-ray spectra with ISGRI data, which are detected up $\sim 150$ keV, higher than $\sim 70$ keV of HXD.
The compTB model could also make a good fit to the joint {\it Suzaku} XIS and IBIS-ISGRI spectra, though giving slightly worse reduced $\chi^2$ than former results of {\it Suzaku} data.
The degeneracy of $\delta$ and $kT_e$ is partly removed, and a set of parameters with apparently better reduced $\chi^2$ can be found.
The best fit parameters are shown in Table~\ref{fit_si},  and the corresponding unfolded spectra are presented in Figure~\ref{bmc_i}.
The fitting results of $\delta$ and $kT_e$ for the three sources do not differ much from each other.
The electron temperature $kT_e$ is in the range of 2.11--3.48 keV, a little higher than the blackbody temperature $kT_s$.
While the values of $\delta$ vary from 31.8 to 51.8, showing that the BMC process would be dominated over the TC process.

\begin{table}
\bc
\begin{minipage}[]{100mm}
\caption[]{The fitting results of {\it Suzaku} and INTEGRAL data with compTB model.  
\label{fit_si}}
\end{minipage}
 \begin{tabular}{cccc}
  \hline\noalign{\smallskip}
 & 1RXS J1708-40 & 1E 1547-54 & SGR 1806-20 \\
  \hline\noalign{\smallskip}
$N_{\rm H} (10^{22} {\rm cm}^{-2})$ & $1.52 \pm 0.01$ & $3.40 \pm 0.03$ & $7.89 \pm 0.15$\\
$kT_{\rm s}$ (keV) & $1.16 \pm 0.01$ & $1.14 \pm 0.03$ & $0.89 \pm 0.02$ \\ 
$\gamma$ & $0.12 \pm 0.01$ & $0.82 \pm 0.02$ & $0.17 \pm 0.01$ \\ 
$\alpha$ & $0.43 \pm 0.01$ & $0.72 \pm 0.02$ & $0.67 \pm 0.01$ \\
$\delta$ & $50.8 \pm 1.7$ & $36.5_{-5.6}^{+29.6}$ & $31.8_{-2.0}^{+2.4}$ \\
$kT_{\rm e}$ (keV) & $2.11 \pm 0.06$ & $3.78_{-0.48}^{+0.81}$ & $3.48_{-0.18}^{+0.22}$ \\ 
log(A) & $-1.54 \pm 0.01$ & $-0.18 \pm 0.02$ & $-0.43 \pm 0.03$ \\ 
$\chi_r^2$ (d.o.f.) & 1.14 (173) & 1.01 (94) & 0.92 (86) \\ 
  \noalign{\smallskip}\hline
\end{tabular}
\ec
\end{table}

\begin{figure}
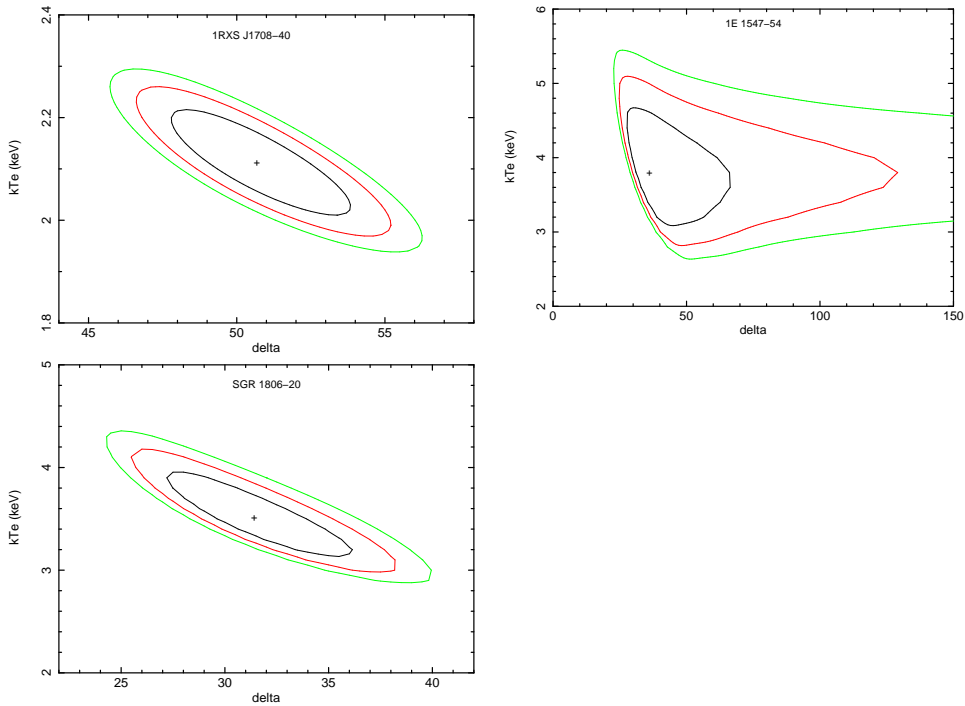

   \includegraphics[height=0.45\textwidth, angle=-90]{1708con.eps}
   \includegraphics[height=0.45\textwidth, angle=-90]{1547con.eps}\\
   \includegraphics[height=0.45\textwidth, angle=-90]{1806con.eps}

   \caption{Error contours of $\delta$(delta) and kTe, for compTB fitting to {\it Suzaku} and INTEGRAL data of 1RXS J1708-40, 1E 1547-54 and SGR 1806-20. The sign `+' indicates parameter values with the minimum $\chi^2$. 1$\sigma$, 2$\sigma$ and 3$\sigma$ contours are in black, red and green respectively.} 
   \label{con}
   \end{figure}

We also draw error contours for different values of $\delta$ and $kT_e$ to explore the reliability of fitting results, as shown in Figure~\ref{con}.
For 1RXS J1708-40, there is a satisfying constraint for the two parameters; while the parameters of 1E 1547-54 are not well constrained, as a result of the relatively short exposure time of INTEGRAL data utilized; the situation of SGR 1806-20 is similar to that of 1RXS J1708-40.

%
%
%
%
%
%

\section{HXMT simulation}

\begin{figure}
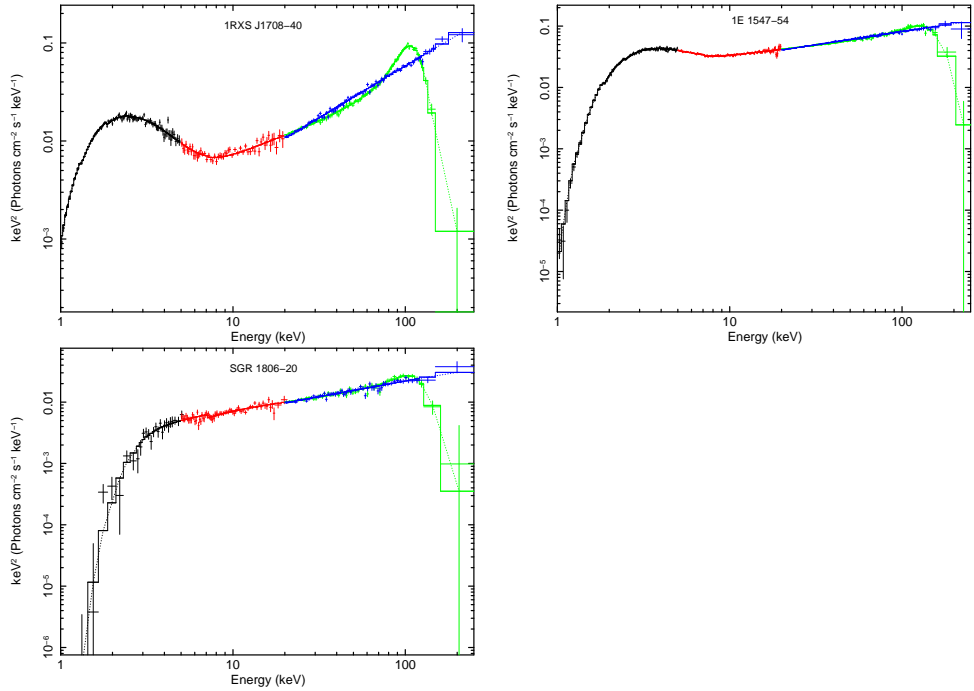

   \includegraphics[height=0.45\textwidth, angle=-90]{1708spe_h.eps}
   \includegraphics[height=0.45\textwidth, angle=-90]{1547spe_h.eps}\\
   \includegraphics[height=0.45\textwidth, angle=-90]{1806spe_h.eps}

   \caption{HXMT simulated spectra of 1RXS J1708-40, 1E 1547-54 and SGR 1806-20. The low (in black) and middle (in red) energy bands are based on parameters of compTB model, while the high energy spectra are simulated for both the compTB and power-law model, shown in green and blue respectively. } 
   \label{spe_h}
   \end{figure}

Although IBIS-ISGRI covers the 20--500 keV energy band, data points above $\sim$ 150 keV have low S/N, and the spectra are insufficient to differentiate models.
The first space telescope of China, HXMT, is to be launched in late 2014 or early 2015.
%
HXMT is a collimated hard X-ray telescope based on the direct demodulation method and NaI(Tl)/CsI(Na) phoswich detecting techniques. 
The payload consists of three telescopes; they work in the low, middle and high energy range respectively, covering the 1--250 keV energy band. 
Among them the high energy telescope, sensitive between 20 and 250 keV, has a large collecting area of 5000 cm$^2$ and hence high sensitivity.
Based on the fitting results of compTB model in Table~\ref{fit_si}, we simulate spectra with HXMT response matrix and background file, shown by green lines in Figure~\ref{spe_h}.
With an exposure time of 1 Ms, obvious cutoff around 200 keV can be seen for the three sources.

\begin{figure}
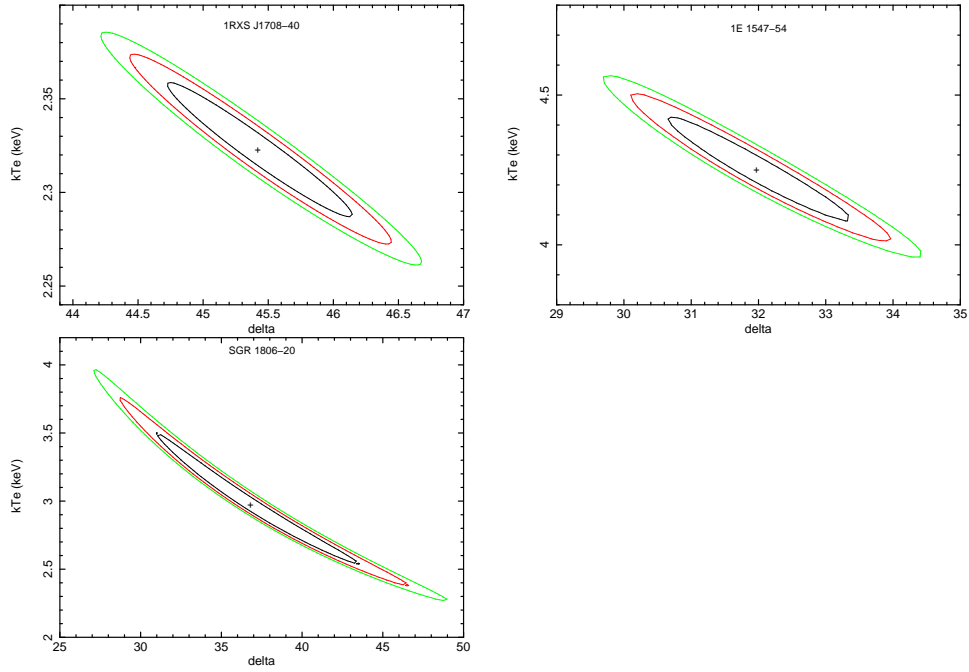

   \includegraphics[height=0.45\textwidth, angle=-90]{1708con_h.eps}
   \includegraphics[height=0.45\textwidth, angle=-90]{1547con_h.eps}\\
   \includegraphics[height=0.45\textwidth, angle=-90]{1806con_h.eps}

   \caption{Error contours of $\delta$(delta) and kTe, for compTB fitting to HXMT simulated spectra of 1RXS J1708-40, 1E 1547-54 and SGR 1806-20. The sign `+' indicates parameter values with the minimum $\chi^2$. 1$\sigma$, 2$\sigma$ and 3$\sigma$ contours are in black, red and green respectively.} 
   \label{con_h}
   \end{figure}

The simulated spectra are also fitted with compTB model to examine the improvement of parameter constraints, and the fitting results are listed in Table~\ref{fit_h}.
The errors of $\delta$ and $kT_e$ are much smaller than the results of {\it Suzaku} and INTEGRAL data, while the constraints of other parameters are similar.
Error contours of $\delta$ and $kT_e$ are also drawn in Figure~\ref{con_h}, which exhibit better constraints than those in Figure~\ref{con}.

\begin{table}
\bc
\begin{minipage}[]{100mm}
\caption[]{The fitting results of HXMT siumlated data with compTB model. 
\label{fit_h}}
\end{minipage}
 \begin{tabular}{cccc}
  \hline\noalign{\smallskip}
 & 1RXS J1708-40 & 1E 1547-54 & SGR 1806-20 \\
  \hline\noalign{\smallskip}
$N_{\rm H} (10^{22} {\rm cm}^{-2})$ & $1.51 \pm 0.01 $ & $3.33 \pm 0.01$ & $8.60_{-0.33}^{+0.35}$\\
$kT_{\rm s}$ (keV) & $1.17_{-0.01}^{+0.05}$ & $1.11 \pm 0.01 $ & $0.86 \pm 0.03$ \\ 
$\gamma$ & $0.13 \pm 0.01 $ & $0.94 \pm 0.01$ & $0.21 \pm 0.02$ \\ 
$\alpha$ & $0.43 \pm 0.01$ & $0.72 \pm 0.01$ & $0.68 \pm 0.01 $ \\
$\delta$ & $45.4 \pm 0.2$ & $32.0 \pm 0.4$ & $36.9 \pm 0.7$ \\
$kT_{\rm e}$ (keV) & $2.32 \pm 0.01$ & $4.25 \pm 0.05$ & $2.97_{-0.05}^{+0.06}$ \\ 
log(A) & $-1.53 \pm 0.01 $ & $-0.27 \pm 0.01$ & $-0.31_{-0.05}^{+0.06}$ \\ 
$\chi_r^2$ (d.o.f.) & 0.95 (220) & 1.15 (337) & 1.03 (146) \\ 
  \noalign{\smallskip}\hline
\end{tabular}
\ec
\end{table}

In the frame of magnetars, the hard X-ray spectrum is generally expected to be power-law with different cutoff properties.
The quantum electrodynamics model by ~\citet{Heyl05} predicts a cutoff energy far above 1 MeV and high flux in $\gamma$-ray band.
However, the $\gamma$-ray flux of 4U 0142+61 is not detected by Fermi-LAT~\citep{Sasmaz10}.
The bremsstrahlung model by \citet{Beloborodov07} might have cutoff at a few hundred keV, but the emerging spectrum will have a photon index $\Gamma \sim 1$ below the break energy for all sources, which is inconsistent with current observations.
The resonant inverse Compton scattering model~\citep{Baring07} also predicts a power-law with cutoff energy higher than 1 MeV.
%
%
%
These hard X-ray models in the magnetar frame can thus be represented by a power-law without cutoff below 250 keV.

According to the parameters of power-law fitting to ISGRI data in Table~\ref{tvar}, we also simulate the HXMT spectra of power-law model, shown by blue lines in Figure~\ref{spe_h}.
Comparing the HXMT simulations of power-law and BMC model, there are obvious discrepancies above $\sim 100$ keV, since the spectra of BMC model could have cutoff around 200 keV; below $\sim 100$ keV, spectra of the two models are roughly similar, but there are also differences in detail.
The relative low cutoff energy of BMC model results from the low electron temperature, different from ultra-relativistic electrons in other models.
%
The quality of INTEGRAL spectra is not able to discriminate these differences, but future HXMT observations should have spectral resolution and S/N high enough to distinguish between those models.
Besides, as we still lack observational data in the 10--20 keV range, the complete 1--250 keV HXMT spectra could provide more information and put better constraints on theoretical models.

\section{conclusions}

Whether AXPs/SGRs are magnetars or quark star/fallback disk systems remains a problem to be settled.
We study the soft and hard X-ray spectra of four AXPs/SGRs with {\it Suzaku} and INTEGRAL observations.
The broad-band {\it Suzaku} spectra could be well reproduced by BMC process, 
%
%
and BMC model could also fit the combined {\it Suzaku} and INTEGRAL spectra, with parameters better constrained.
Thus fallback disk system could be compatible with X-ray emission of AXPs/SGRs, implying that the existence of accretion flow is possible.
In addition, HXMT simulated spectra of BMC model exhibit cutoff around 200 keV, showing significant discrepancy from power-law spectra. 
%
%
We can expect to distinguish BMC model from other hard X-ray models in the magnetar frame by future Chinese HXMT observations, and further understand the nature of AXPs/SGRs.

%


\begin{acknowledgements}

We would like to thank useful discussions at pulsar group at PKU.
This research has made use of data and software obtained from NASA's High Energy Astrophysics Science Archive Research Center (HEASARC), a service of Goddard Space Flight Center and the Smithsonian Astrophysical Observatory.
%
This work is supported by the National Basic Research Program of China (2012CB821800), the National Natural Science Foundation of China (11103019,11103021,11225314, 11203018), the National Fund for Fostering Talents of Basic Science and XTP project XDA04060604.

\end{acknowledgements}
  
\bibliographystyle{raa}
\bibliography{ref}

\end{document}